\documentclass{article}

\usepackage{amssymb}
\usepackage{citesort}
\usepackage{amsmath}

\def\ud{\mathrm{d}}

\begin{document}
\vspace*{1.0cm}
\noindent
{\bf
{\large
\begin{center}
On the uniqueness of paths for spin-0 and spin-1 quantum mechanics
\end{center}
}
}

\vspace*{.5cm}
\begin{center}
W.\ Struyve, W.\ De Baere, J.\ De Neve and S.\ De Weirdt
\end{center}

\begin{center}
Laboratory for Theoretical Physics\\
Unit for Subatomic and Radiation Physics\\
Proeftuinstraat 86, B--9000 Ghent, Belgium\\
E--mail: ward.struyve@ugent.be
\end{center}

\begin{abstract}
\noindent
The uniqueness of the Bohmian particle interpretation of the Kemmer equation, which describes massive spin-0 and spin-1 particles, is discussed. Recently the same problem for spin-$\frac{1}{2}$ was dealt with by Holland. It appears that the uniqueness of boson paths can be enforced under well determined conditions. This in turn fixes the nonrelativistic particle equations of the nonrelativistic Schr\"odinger equation, which appear to correspond with the original definitions given by de Broglie and Bohm only in the spin-0 case. Similar to the spin-$\frac{1}{2}$ case, there appears an additional spin-dependent term in the guidance equation in the spin-1 case. We also discuss the ambiguity associated with the introduction of an electromagnetic coupling in the Kemmer theory. We argue that when the minimal coupling is correctly introduced, then the current constructed from the energy--momentum tensor is no longer conserved. Hence this current can not serve as a particle probability four-vector.
\end{abstract}

\section{Introduction}
In a recent paper \cite{holland99}, Holland demonstrated that the multi-particle Dirac current is unique under the constraints that the current is a Lorentz covariant quantity and that it reproduces the empirically well-established spin-$\frac{1}{2}$ particle probability density. As a result, the Bohmian laws of motion for spin-$\frac{1}{2}$ particles, which arise when the Dirac current is interpreted as a real particle current, are uniquely defined. By taking the nonrelativistic limit, the uniqueness of the Dirac current implied a unique Schr\"odinger current corresponding to the nonrelativistic many-particle Schr\"odinger equation. It appeared that this current differs from the conventional Schr\"odinger current by an additional spin-dependent term. As a result the corresponding Bohmian particle laws generally differ from those originally presented by de Broglie and Bohm for spin-$\frac{1}{2}$ particles. Recently, the implications of this additional spin-dependent term have been studied for the particle paths in the two slit experiment \cite{holland031}, and for hydrogen eigenstates \cite{colijn02} and transitions between them \cite{colijn03}.

As mentioned by Holland, the original equations of de Broglie and Bohm should be valid only for massive spin-0 particles. In this Letter we address this problem of uniqueness for the Bohmian trajectories in the Klein--Gordon and Proca theory. We depart hereby from the Kemmer theory, which recasts the massive Klein--Gordon and Proca formalism in an elegant way, formally resembling the Dirac theory \cite{kemmer39}. Because the spin-0 and spin-1 charge currents do not admit a particle interpretation, we start with the energy--momentum tensor from which a conserved, future-causal four-vector can be constructed with positive time component \cite{ghose93,ghose94,ghose96,ghose01}. It appears that Holland's method of imposing uniqueness for spin-$\frac{1}{2}$ particles cannot be extended to the case of spin-0 and spin-1. I.e.\ the condition that the Kemmer energy--momentum tensor correctly reproduces the observable energy density is not sufficient to impose its uniqueness. Only under further restrictions we can establish a unique energy--momentum tensor. In this paper we adopt an additional restriction, initially explored by Fock \cite{fock64} and Holland \cite{holland032}. Under this additional condition we show that the Bohmian trajectories, in both the single-particle and the multi-particle case, are uniquely defined. By taking the nonrelativistic limit it is shown that, in the spin-0 case, the Kemmer energy--momentum current results in the conventional current corresponding to the Schr\"odinger equation. For a nonrelativistic spin-1 eigenstate (which obeys the Schr\"odinger equation) an additional spin-dependent term appears in the Schr\"odinger current. As a result the Bohmian particle laws, which are derived from these currents, result in the original laws given by de Broglie and Bohm only in the spin-0 case. In the spin-1 case there appears an additional spin-dependent term in the guidance equation. It is interesting to note that in the nonrelativistic limit, the Kemmer charge current, although it can not be interpreted as a particle current, leads to the same nonrelativistic currents as those derived from the energy--momentum tensor. 

We also analyze the electromagnetic coupling in the Kemmer theory. There seem to be two inequivalent ways of introducing an electromagnetic field via minimal coupling. One can do it at the level of the covariant form or at the Hamiltonian form of the Kemmer equations. It was recently stated that the two ways are equally applicable because the difference would disappear when the minimal coupled Kemmer theory is reduced to its physical components \cite{nowakowski98,lunardi00}. However we argue that in general the only correct way to introduce minimal coupling is at the level of the covariant form of the Kemmer equations. An important consequence of introducing minimal coupling at this level is that the Kemmer energy--momentum tensor is no longer conserved. Hence the energy--momentum vector constructed from this tensor cannot represent a conserved particle current.

\section{Uniqueness of paths in the one-particle  spin-0 and spin-1 case} 
To describe spin-0 and spin-1 relativistic particles with mass $m$, we can start from the Kemmer equations \cite{kemmer39} (with units in which $c=\hbar=1$)
\begin{equation}
(i\beta^{\mu} \partial_{\mu} - m )\psi = 0, \quad i\partial^{\mu}  {\bar \psi} \beta_{\mu} + m {\bar \psi} = 0,
\label{1.1}
\end{equation}
where ${\bar \psi} =\psi^{\dagger} \eta_0 $ with $\eta_0 = 2\beta^2_0 - 1 $, and the Kemmer-Duffin-Petiau matrices $\beta_\mu$ satisfy the commutation relations
\begin{equation}
\beta_{\mu}\beta_{\nu}\beta_{\lambda} + \beta_{\lambda}\beta_{\nu}\beta_{\mu} = \beta_{\mu}g_{\nu \lambda} + \beta_{\lambda} g_{\nu \mu}.
\label{2}
\end{equation}
There are three inequivalent irreducible representations of the $\beta_{\mu}$, one is $10 \times 10$ which amounts to the Proca equations and describes spin-1 bosons, another one is $5 \times 5$ which amounts to the Klein--Gordon equation and describes spin-0 bosons and the third one is the trivial $1 \times 1$ representation. The Kemmer equations can be derived from the Lagrangian density 
\begin{equation}
\mathcal{L}_K = \frac{i}{2} ({\bar \psi} \beta_{\mu}\partial^{\mu} \psi  - \partial^{\mu}  {\bar \psi} \beta_{\mu} \psi ) -m{\bar \psi}\psi.
\label{3}
\end{equation}
The corresponding conserved symmetrized energy--momentum tensor is
\begin{equation}
\Theta^{\mu \nu}_K =  {\bar \psi} (\beta^{\mu}\beta^{\nu} +\beta^{\nu}\beta^{\mu} - g^{\mu \nu})\psi.
\label{6.1}
\end{equation}
For an observer with constant four-velocity $a^\mu$ the energy density is given by $\Theta^{\mu \nu}_K a_\mu a_\nu$. It is then natural to define the conserved energy--momentum vector for the observer with four-velocity $a^\mu$ as 
\begin{equation}
j^{\mu} = \Theta^{\mu \nu}_K a_{\nu},
\label{7}
\end{equation}
see \cite{holland}{\footnote{This way of constructing a conserved current from the energy--momentum tensor is to be distinguished from the approach by Dewdney, Horton and Nesteruk \cite{horton00,horton01} who use a future-causal eigenvector of the energy--momentum tensor instead of the four-velocity of an observer. In this way they are able to give a causal trajectory interpretation to the intrinsicly defined  energy--momentum vector.}}. The current $j^\mu$ is conserved, i.e.\ $\partial_{\mu} j^{\mu} = 0$, and future-causal, i.e.\ $j^0 \geqslant 0$ and $j^{\mu} j_{\mu} \geqslant 0$. The fact that $j^{\mu}$ is future-causal can be derived as follows. Because $\Theta^{00}_K =  \psi^{\dagger} \psi \geqslant 0$ and $\Theta^{0\mu }_K \Theta_{K0\mu } \geqslant 0$ (which can be verified in the explicit representations used in section 3), the vector $\Theta^{0 \mu }_K = \delta^0_\nu \Theta^{\nu \mu }_K $ is future-causal. Because the product of two future-causal vectors is positive, $j^0 = \Theta^{0\mu }_K a_{\mu}\geqslant 0$ for $a^{\mu}$ future-causal. The fact that $j^{\mu} j_{\mu} \geqslant 0$ can be seen if we perform a Lorentz transformation such that $\Lambda^\mu_{\ \nu} a^{\nu} =\delta^\mu_0$ because then $j^{\mu} j_{\mu} =\Theta'^{0\mu }_K \Theta'_{K0\mu }$, with $\Theta'^{\mu_1 \mu_2 }_K = \Lambda^{\mu_1}_{\ \nu_1} \Lambda^{\mu_2}_{\ \nu_2} \Theta^{\nu_1  \nu_2 }_K$ and $\Theta'^{0\mu }_K \Theta'_{K0\mu }$ is positive as it has the same form as the positive quantity $\Theta^{0\mu }_K \Theta_{K0\mu}$ (just replace $\psi({\bf x},t)$ in the energy--momentum tensor by $\psi'({\bf x}',t')$, where the accents refer to quantities in the new Lorentz frame).

These properties of the current $j^\mu$ led Ghose {\em et al.} \cite{ghose93,ghose94,ghose96,ghose01} to interpret it, in the spirit of de Broglie and Bohm, as a particle probability four-vector current. The probability density of the particles in the ensemble is then taken as $j^0$, and the four-velocity field of the particles is then defined as $u^\mu =j^{\mu}/j^\nu j_\nu$. The paths ${\bf x} = {\bf x}(t,{\bf x}_0) $ of the particles can then be found by solving the ``guidance equation''
\begin{equation}
\frac{\ud x^i}{ \ud t} = \frac{u^{i}}{u^0} = \frac{j^i}{j^0}.
\label{8}
\end{equation}
Because the vector $u^\mu$ is timelike one has the relativistic requirement that the Bohmian speeds do not exceed the speed of light, i.e.\ $\big| \ud {\bf x}/ \ud t\big| \leqslant 1$. In the special case that $a^{\mu} = \delta^\mu_0$, i.e.\ for an observer at rest, the probability density is given by the energy density $\Theta^{00}_K= \psi^{\dagger} \psi$ and the guidance equation reads
\begin{equation}
\frac{\ud x^i}{\ud t} = \frac{\Theta^{i0}_K}{\Theta^{00}_K} = \frac{\psi^{\dagger} {\tilde \beta}_i \psi}{\psi^{\dagger} \psi},
\label{9}
\end{equation}
where ${\tilde \beta}_i = \beta_0 \beta_i - \beta_i \beta_0$. Note the formal analogy with the corresponding expressions for the Bohmian particles guided by the Dirac wavefunction \cite{holland,bohm5}. 

The technique of constructing a conserved current from the energy--momentum tensor by contracting it with the four-velocity of an observer was originated by Holland \cite{holland}, who applied it to the case of photons and massive spin-0 particles. However, Holland was reluctant to regard the energy--momentum current as a particle probability current and rather regarded the guidance equation in ({\ref{9}) as the defining formula for the tracks of energy flow. Holland gave a series of arguments to support this view. Most of the arguments were against the notion of a photon as a localized particle and do not apply to massive bosons which can be built up from fermions, for which the particle concept is well accepted. In this paper we will not dwell with the question whether the particle interpretation of the energy--momentum current can be manifestly maintained (in fact this is more an experimental issue, where the experimental results should be correctly interpreted) and we will pursue the suggestion of Ghose {\em et al.} to give a particle interpretation to bosons via the energy--momentum tensor. Nevertheless we want to note that in section 5 we argue that the energy--momentum tensor is in general no longer conserved when interaction with an electromagnetic field is introduced. Hence in the coupled case a particle interpretation of the energy--momentum vector is no longer tenable. 

One can now consider the question whether the trajectories defined by ({\ref 8}) are uniquely defined. A priori there could be other possible guidance laws arising from possible alternative definitions of the energy--momentum tensor. This is because one is always free to add a divergenceless tensor $A^{\mu \nu}$ to the energy--momentum tensor $\Theta^{\mu \nu}_K$. The newly defined tensor 
\begin{equation}
{\bar \Theta}^{\mu \nu}_K = \Theta^{\mu \nu}_K + A^{ \mu \nu}
\label{10}
\end{equation}
would then also imply a conserved current 
\begin{equation}
{\bar j}^{\mu} =  j^{\mu} +  A^{\mu \nu} a_{\nu} 
\label{11}
\end{equation}
which could be interpreted as a particle probability four-vector current. 

A similar problem arises in the Dirac theory. By addition of a divergenceless vector $A^\mu$ to the Dirac current $J^\mu={\bar{\psi} \gamma^\mu \psi}$ one can always construct another conserved current ${\bar J}^{\mu}=J^\mu + A^\mu$, which could also be interpreted as a particle probability current. In \cite{holland99} Holland established the uniqueness of the Dirac current by observing that the spin-$\frac{1}{2}$ particle probability distribution $\psi^{\dagger}\psi$ is empirically well-determined. The requirement that the current ${\bar J}^{\mu}$ must reproduce this probability distribution in every Lorentz frame, which means that ${\bar J}^{0} = J^0 = \psi^{\dagger}\psi$ in every frame, then implies that the additional vector $A^\mu$ must have a zero time component in every frame and hence is identically zero. 

In the Kemmer case we cannot demand that ${\bar j}^{0} =  j^{0}$ in every Lorentz frame, because $j^0$ is in general not an observable distribution. In the Kemmer case, the observable quantity associated with the energy--momentum tensor is the energy density $\Theta^{\mu \nu}_K a_\mu a_\nu$ (regardless whether it is interpreted as a real particle density or rather as an energy density). Hence only in the frame which is at rest relative to the observer, i.e.\ when $a^\mu = \delta^\mu_0$, the time component of the current $j^\mu$ is observable. Because for $a^\mu = \delta^\mu_0$ the time component is the energy density, i.e.\ $j^0 = \Theta^{00}_K=\Theta^{\mu \nu}_K \delta^0_\mu \delta^0_\nu$. 

We can thus restrict the energy--momentum tensor and hence the energy--momentum current for any observer with constant four-velocity, by demanding that it must give the correct energy distribution for these observers. This means that one should have ${\bar \Theta}^{\mu \nu}_K a_\mu a_\nu= \Theta^{\mu \nu}_K a_\mu a_\nu$ for every $a^\mu$ which is constant and future-causal. This implies that $A^{\mu \nu} a_\mu a_\nu = 0$ for every constant future-causal vector $a^\mu$, which in turn implies that $A^{\mu \nu}$ must be antisymmetric. This leaves us with a multitude of possible energy--momentum tensors that give the correct observable energy density and hence with a multitude of possible definitions for the energy--momentum vector for an arbitrary observer.

Remark that the vector $A^{\mu \nu} a_{\nu}$ in ({\ref{11}) is space-like because it is orthogonal to $a^\mu$, which is time-like. This property tends to destroy the overall time-like character of the current ${\bar j}^{\mu}$ and hence the causal character of the Bohmian trajectories. However, there is no clear evidence that ${\bar j}^{\mu}$ becomes space-like at certain space-time configurations where $A^{\mu \nu} a_{\nu}$ is nonzero.

In order to remove the remaining indeterminacy of the current we will have to resort to additional assumptions on the character of $A^{\mu \nu}$.  A possible assumption which we shall adopt in this Letter, is that the energy--momentum tensor is only dependent on the fields and derivatives of them which are of order one less than appears in the field equations. Hence $A^{\mu \nu}$ should only be dependent on the Kemmer wavefunction $\psi$ and its complex conjugate $\psi^*$, and not on their space or time derivatives. This additional condition will restrict the tensor $A^{\mu \nu}$ to be zero. The method of proof was initiated by Fock \cite{fock64} and further elaborated by Holland \cite{holland032} who derived unique expressions for respectively the energy--momentum tensor of the electromagnetic field, and the Dirac and the Klein--Gordon charge currents. As Holland pointed out, allowance of higher order derivatives of the field in the Dirac and Klein--Gordon currents results in a multitude of possible conserved currents. This will also be the case for the Kemmer energy--momentum tensor. 

Under the mentioned assumptions the Kemmer energy--momentum tensor $\Theta^{\mu \nu}_K$ is determined up to a tensor $A^{\mu \nu} = A^{\mu \nu}(\psi,\psi^{\dagger})$ which satisfies $\partial_\mu A^{\mu \nu}=0$ and $A^{\mu \nu} = - A^{\nu \mu}$. Let us now express the conservation of the additional tensor as a constraint of the functional dependence of the tensor on the Kemmer wavefunction 
\begin{equation}
\frac{\partial A^{\mu \nu}}{\partial \psi} \partial_\mu \psi + \partial_\mu \psi^{\dagger} \frac{\partial A^{\mu \nu}}{\partial \psi^{\dagger}} = 0.
\label{11.2}
\end{equation}
Due to the antisymmetry of $A^{\mu \nu}$ one has for $\nu = 0$ 
\begin{equation}
\frac{\partial A^{i 0}}{\partial \psi} \partial_i \psi + \partial_i \psi^{\dagger} \frac{\partial A^{i 0}}{\partial \psi^{\dagger}} = 0.
\label{11.3}
\end{equation}
Because $A^{\mu \nu}$ is independent of derivatives of the Kemmer wavefunction and because $\partial_i \psi$ and $\partial_i \psi^{\dagger}$ can be treated as independent arbitrary functions, one has
\begin{equation}
\frac{\partial A^{i 0}}{\partial \psi} = \frac{\partial A^{i 0}}{\partial \psi^{\dagger}} = 0.
\label{11.31}
\end{equation}
This implies that the components $A^{i 0}$ are constant. Under the natural assumption that the energy--momentum tensor vanishes when the wavefunction is zero, the components $A^{i 0}$ are zero. Because these relations apply to each Lorentz frame, all the components of the tensor $A^{\mu \nu}$ are zero. 

This establishes that the Kemmer energy--momentum tensor is unique under the conditions (a) that it is conserved, (b) that it gives the conventional expression for the energy density for any observer with constant four-velocity (which is empirically verifiable) and (c) that it is only dependent on the Kemmer wavefunction and its complex conjugate. If we relinquish condition (c) and allow derivatives of the Kemmer wavefunction in the energy--momentum tensor, a multitude of tensors obeying conditions (a) and (b) are possible. For example if we take a totally antisymmetric tensor $\chi^{\sigma \mu \nu }$ then the tensor $\partial_\sigma \chi^{\sigma \mu \nu}$ is identically conserved and antisymmetric. 

\section{Nonrelativistic limit of the Kemmer equations}
Before we take the nonrelativistic limit, we write the Kemmer equations in an equivalent form, which is more convenient to recover the Klein--Gordon equation in the spin-0 representation, and the Proca equations in the spin-1 representation. The Kemmer equation ({\ref{1.1}}) is equivalent with the two equations 
\begin{eqnarray}
i \partial_{0} \psi &=& (-i{\tilde \beta}_{i} \partial_{i} + m\beta_{0}   ) \psi \label{13.1}\\
i \beta_{i} \beta^2_{0}\partial_{i}\psi &=& m (1 -  \beta^2_{0}) \psi.
\label{13.2} 
\end{eqnarray}
The first equation is a Schr\"odinger-like equation and the second equation has to be regarded as an additional constraint on the wavefunction $\psi$. We will now consider the two cases, spin-0 and spin-1, separately.

\subsection{The spin-0 case}
On using explicit matrix representations (i.e.\ the ones used in \cite{ghose96}, where the $\beta_i$ correspond to $-i\beta_i$ in \cite{kemmer39}), the second equation ({\ref{13.2}}) implies a Kemmer wavefunction $\psi$ with the following five components: $\omega$, $\partial_{1} \phi$, $\partial_{2} \phi$, $\partial_{3} \phi$, $m\phi$ with $\omega$ and $\phi$ for the moment arbitrary scalar wavefunctions. Equation ({\ref{13.1}}) further determines $\omega$ as $\omega= \partial_{0} \phi$. In this way the wavefunction $\psi$ can be conveniently written as 
\begin{equation}
\psi = (\partial_\mu \phi,m \phi)^T.
\label{13.21}
\end{equation}
Equation ({\ref{13.1}}) then reduces to the massive Klein--Gordon equation for $\phi$, which is the remaining physical component of $\psi$ 
\begin{equation}
\square \phi + m^2 \phi=0.
\label{13.3}
\end{equation}
Substitution of the wavefunction ({\ref{13.21}}) into the (uniquely defined) Kemmer energy--momentum tensor ({\ref{6.1}}) renders the Klein--Gordon energy--momentum tensor for $\phi$
\begin{equation}
\Theta^{\mu \nu}_{KG} = \partial^{\mu} \phi \partial^{\nu} \phi^* + \partial^{\mu} \phi^* \partial^{\nu} \phi - g^{\mu \nu} (\partial_{\alpha} \phi \partial^{\alpha} \phi^* - m^2 \phi^* \phi),
\label{15}
\end{equation}
which can be derived from the Klein--Gordon Lagrangian 
\begin{equation}
\mathcal{L}_{KG}  = \partial_{\alpha} \phi \partial^{\alpha} \phi^* - m^2 \phi^* \phi.
\label{16}
\end{equation}
Because $\Theta^{\mu \nu}_{K}$ equals $\Theta^{\mu \nu}_{KG}$ when relation ({\ref{13.21}}) is taken into acount, the Bohmian trajectory laws for spin-0 particles result in the trajectory laws for the Klein--Gordon case originally proposed by Holland. 

To take the nonrelativistic limit of the Klein--Gordon equation, one substitutes $\phi = e^{-imt} \psi'/\sqrt{2}m$, where the energy of the wavefunction $\psi'$ is much smaller than the rest energy $m$. As is well known, the substitution of $\phi$ in the Klein--Gordon equation leads to the Schr\"odinger equation for $\psi'$. The nonrelativistic limit of the Klein--Gordon energy--momentum tensor becomes
\begin{eqnarray}
\Theta^{00}_{KG} &=& |\psi'|^2,\quad \Theta^{i0}_{KG} = \frac{1}{m} \textrm{Im} \big( \psi'^* \partial_i \psi' \big)  \nonumber\\
\Theta^{ij}_{KG} &=& \frac{1}{m^2} \textrm{Re}(\partial_i \psi' \partial_j \psi'^*) + \delta_{ij} \bigg(\frac{\textrm{Im}(\psi'\partial_0\psi'^* ) }{m} - \frac{\partial_k \psi' \partial_k \psi'^*}{2m^2} \bigg).
\label{17}
\end{eqnarray}
Because one has $\Theta^{00}_{KG} \gg \Theta^{i0}_{KG} \sim p/m \gg \Theta^{ij}_{KG}\sim p^2/m^2$ and $a^0 \gg |{\bf{a}}|$, the nonrelativistic limit of current $j^\mu= \Theta^{\mu \nu}_{KG} a_\nu$ becomes the conventional Schr\"odinger current
\begin{equation}
j^0  \! = \Theta^{00}_{KG} = |\psi'|^2, \quad j^i = \Theta^{i0}_{KG} = \frac{1}{m} \textrm{Im} ( \psi'^*  \nabla_i \psi' ).
\label{17.01}
\end{equation}
The corresponding Bohmian spin-0 particle density and quidance equation then become 
\begin{equation}
\rho = |\psi'|^2, \quad  \frac{\ud {\bf x}}{ \ud t} = \frac{\textrm{Im} ( \psi'^* {\boldsymbol \nabla} \psi' )}{m |\psi'|^2} .
\label{17.1}
\end{equation}
These were the equations originally used by de Broglie and Bohm to give a particle interpretation to the Schr\"odinger equation \cite{holland,bohm5}. 

This establishes that the Bohmian equations for spin-0 particles, i.e.\ the guidance equation and the form of the probability density, derived as the nonrelativistic limit of the relativisic Kemmer theory, are unique under the conditions stated at the end of section 2 and correspond to the original definitions.

\subsection{The spin-1 case}
The discussion of the spin-1 case proceeds analogously. If we take the ten components of the Kemmer wavefunction as: 
\begin{equation}
\psi = (-{\bf E}, {\bf B},m{\bf A}, -mA_0)^T,
\label{18.1}
\end{equation}
then equation ({\ref{13.2}}) implies the following relations
\begin{equation}
 {\boldsymbol \nabla} \cdot {\bf E} = - m^2 A_0, \qquad {\bf B} =  {\boldsymbol \nabla} \times {\bf A}.
\label{19}
\end{equation}
The equation ({\ref{13.1}}) leads to the relations
\begin{eqnarray}
\partial_0 {\bf E}  &=&  {\boldsymbol \nabla} \times {\bf B}+ m^2 {\bf A}, \quad \partial_0 {\bf B} = -  {\boldsymbol \nabla} \times {\bf E} \nonumber\\
 {\bf E} &=& - {\boldsymbol \nabla} A_0 - \partial_0 {\bf A} , \quad \partial_\mu A^\mu = 0.
\label{20}
\end{eqnarray}
If one takes instead as definitions
\begin{equation}
{\bf B} =  {\boldsymbol \nabla} \times {\bf A}, \qquad {\bf E} = - {\boldsymbol \nabla} A_0 - \partial_0 {\bf A}, \qquad G^{\mu \nu} = \partial^{\mu} A^{\nu} - \partial^{\nu} A^{\mu}
\label{21}
\end{equation}
then the remaining relations in ({\ref{19}}) and ({\ref{20}}) can be shown to arise from the Proca equations\begin{equation}
\partial_\mu G^{\mu \nu} = - m^2 A^{\nu}
\label{22}
\end{equation}
for the field $A^\mu$. The Proca equations are in turn derivable from the Proca Lagrangian
\begin{equation}
\mathcal{L}_P  = - \frac{1}{2} G^*_{\mu \nu}G^{\mu \nu} + m^2 A^*_{\mu}A^{\mu}.
\label{23}
\end{equation}
The symmetrized Proca energy--momentum tensor reads   
\begin{equation}
\Theta^{\mu \nu}_P = -G^{* \mu \alpha} G^{\nu}_{\ \alpha} - G^{ \mu \alpha} G^{*\nu}_{\ \ \alpha} - g^{\mu \nu} \mathcal{L}_P + m^2 (A^\mu A^{* \nu} + A^{* \mu} A^\nu ).
\label{24}
\end{equation}
and equals the Kemmer energy--momentum tensor when the relations ({\ref{18.1}}),({\ref{19}}) and ({\ref{20}}) are taken into account.

To take the nonrelativistic limit we write the Proca equations as a Klein--Gordon equation for each component of the field $A^\mu$
\begin{equation}
\square A^\mu + m^2 A^\mu=0
\label{24.0}
\end{equation}
with subsidiary condition
\begin{equation}
\partial_\mu A^\mu =0.
\label{24.01}
\end{equation}
We again separate the rest energy by putting $A^\mu = e^{-imt}\phi^\mu /\sqrt{2} m$. After substitution, the condition $\partial_\mu A^\mu =0$ yields $\phi_0 = \partial_i \phi_i/im$, which implies that we can take $\phi_0$ as the small component of the wavefunction $\phi^\mu$. The nonrelativistic limit of equation ({\ref{24.0}}) results in the Schr\"odinger equation for each component $\phi^\mu$. If we define the wavefunction $\Phi$ with components $\phi_1$, $\phi_2$ and $\phi_3$, then the nonrelativistic limit of the Proca equations can be written as
\begin{equation}
i\frac{\partial \Phi}{\partial t}  = -\frac{\nabla^2 \Phi}{2m} .
\label{24.02}
\end{equation}  
Similarly as in the spin-0 case the nonrelativistic limit of the spin-1 current $j^{\mu} = \Theta^{\mu \nu}_P a_{\nu}$ becomes 
\begin{equation}
j^0  \! = \Theta^{00}_{P} = \Phi^{\dagger} \Phi, \quad j^i = \Theta^{i0}_{P} = \frac{1}{m} \textrm{Im} \big(\Phi^{\dagger}  \nabla_i \Phi  \big) +  \frac{1}{2m} {\boldsymbol \nabla} \times (\Phi^{\dagger} {\bf  {\hat  S}} \Phi ), 
\label{24.03100101}
\end{equation}
where ${\bf {\hat  S}}$ is the spin vector, with components $( {\hat S}_j)_{ik} = i\epsilon_{ijk}$. The current resembles the conventional current corresponding to the nonrelativistic spin-1 equation ({\ref{24.02}}). The difference lies in the presence of the divergenceless spin term.

The corresponding Bohmian equations are given by
\begin{equation}
\rho = \Phi^{\dagger} \Phi, \quad \frac{\ud {\bf x}}{ \ud t} =  \frac{\textrm{Im} ( \Phi^{\dagger}  {\boldsymbol \nabla}  \Phi)}{m \Phi^{\dagger} \Phi} + \frac{{\boldsymbol \nabla} \times (\Phi^{\dagger} {\bf  {\hat  S}} \Phi )}{2m\Phi^{\dagger} \Phi}.
\label{24.05}
\end{equation}
If $\Phi$ is a spin eigenstate, i.e. $\Phi(x) = \psi'(x) \epsilon, \quad \epsilon^{\dagger} \epsilon = 1$, then the Bohmian equations become 
\begin{equation}
\rho = |\psi'|^2, \quad \frac{\ud {\bf x}}{ \ud t} =  \frac{\textrm{Im} \big( \psi'^* {\boldsymbol \nabla}  \psi' \big)}{m|\psi'|^2} + \frac{{\boldsymbol \nabla} \times (|\psi'|^2 \epsilon^{\dagger} {\bf  {\hat  S}} \epsilon) }{2m|\psi'|^2}.
\label{25.05}
\end{equation}
It follows that, compared to the original de Broglie--Bohm guidance equation for the Schr\"odinger equation, there appears an additional spin-dependent contribution when the particle has spin 1. This is similar to the case of spin-$\frac{1}{2}$, where the guidance equation also contains a spin contribution \cite{bohm5,holland99}. In fact the spin term is formally equal to the spin term in the case of spin-$\frac{1}{2}$. If the unit spin vector $\epsilon^{\dagger} {\bf  {\hat  S}} \epsilon$ equals the unit spin-$\frac{1}{2}$ vector $\chi^{\dagger} {\boldsymbol  {\hat  \sigma}} \chi$ (with $\chi$ a configuration independent Pauli spinor and $ {\hat \sigma}_i$ the Pauli matrices), the spin-1 and spin-$\frac{1}{2}$ trajectories are the same. For example the spin-$\frac{1}{2}$ trajectories for the two-slit experiment \cite{holland031} apply for the spin-1 case as well. 

\subsection{Nonrelativistic limit of the charge current}
In the same way as done in section 2, one can prove that the conserved Kemmer charge current $s^{\mu}_K = {\bar \psi} \beta^{\mu} \psi/m$, which transforms as a four-vector under Lorentz transformations, is the only four-vector for which the time component equals the charge density in every Lorentz frame. However, because the charge density is not always positive we cannot interpret the charge current as a particle current. When the Kemmer wavefunction is reduced to its physical components, using ({\ref{13.21}}) in the  spin-0 case and ({\ref{18.1}}) in the spin-1 case, the Kemmer charge current becomes respectively the Klein--Gordon charge current
\begin{equation}
s^{\mu}_{KG} = i\big(\phi^* \partial^{\mu}\phi -\phi \partial^{\mu}\phi^*  \big) 
\label{25.06}
\end{equation}
and the Proca charge current
\begin{equation}
s^{\mu}_{P} = i\big(A^{\nu} G^*_{\mu \nu} - G_{\mu \nu} A^{*\nu} \big).
\label{25.07}
\end{equation}
It is now interesting to note that in the nonrelativistic limit, the Klein--Gordon charge current results in the conventional Schr\"odinger current ({\ref{17.01}}) corresponding to the nonrelativistic Schr\"odinger equation. The Proca charge current results in the current ({\ref{24.03100101}}), which was derived as the nonrelativistic limit of the Proca energy--momentum tensor. We can conclude that in both the spin-0 and spin-1 case the nonrelativistic limit of the energy--momentum tensor and the charge current lead to the same unique Schr\"odinger current corresponding to the nonrelativistic Schr\"odinger equation. This is not the case in the spin-$\frac{1}{2}$ Dirac theory. The components of the symmetrized Dirac energy--momentum tensor $T^{\mu 0}$, lead to the conventional Pauli current in the nonrelativistic limit. For a spin eigenstate the current becomes the conventional Schr\"odinger current. Whereas the nonrelativistic limit of the Dirac charge current implies an additional spin contribution.

\section{The many-particle Kemmer formalism}
The extension of the single-particle Kemmer theory to a $N$-particle system is made as follows. One defines a wavefunction $\psi^{\alpha_1 \dots \alpha_N }({\bf x}_1, \dots,{\bf x}_N, t)$ with $N$ spin indices and operators $\beta^{(\alpha)}_{\mu}$ which operate only on the spin index $\alpha$, belonging to the $\alpha^{th}$ particle. The Schr\"odinger form of the Kemmer equations for the $N$-particle system then reads
\begin{eqnarray}
i \partial_{0} \psi &=& \sum^N_{\alpha = 1} \Big(-i {\tilde \beta}^{(\alpha)}_{i} \partial^{(\alpha)}_{i} + \beta_{0}^{(\alpha)} m_{\alpha}   \Big) \psi \label{26.1}\\
i \beta^{(\alpha)}_{i} \big(\beta^{(\alpha)}_{0} \big)^2 \partial^{(\alpha)}_{i}\psi &=&  m_{\alpha} \bigg[ 1 -  \big(\beta^{(\alpha)}_{0}\big)^2 \bigg] \psi,
\label{26.2} 
\end{eqnarray}
where $\partial^{(\alpha)}_{i} = \frac{\partial}{\partial({\bf x}_{\alpha})_i}$ and $m_{\alpha}$ is the mass of the $\alpha^{th}$ particle. In order to construct the multi-particle generalization of the one-particle energy--momentum tensor ({\ref{6.1}}) we first define the following operator
 \begin{equation}
\Gamma^{(\alpha)}_{\mu \nu} =\eta^{(\alpha)}_0 (\beta^{(\alpha)}_{\mu}\beta^{(\alpha)}_{\nu} +\beta^{(\alpha)}_{\nu}\beta^{(\alpha)}_{\mu} - g_{\mu \nu}).
\label{26.3}
\end{equation}
The multi-particle energy--momentum tensor of rank $2N$ then reads (see also \cite{ghose01})
\begin{equation}
\Theta_{\mu_1 \dots \mu_{2N}} = \psi^{\dagger}\Gamma^{(1)}_{\mu_1 \mu_2} \dots \Gamma^{(N)}_{\mu_{2N-1} \mu_{2N}}\psi.
\label{27}
\end{equation}
The energy--momentum tensor satisfies the conservation equation
\begin{equation}
\partial_0 \Theta^{0_1 \nu_1 \dots 0_N \nu_{N}} + \sum^N_{\alpha = 1}  \partial^{(\alpha)}_{i} \Theta^{0_1 \nu_1 \dots i_{\alpha} \nu_{\alpha} \dots 0_N \nu_N} = 0.
\label{27.1}
\end{equation}

From the tensor $\Theta_{\mu_1 \dots \mu_{2N}}$ one can construct the following tensor of rank $N$
\begin{equation}
j_{\mu_1 \dots \mu_N} = \Theta_{\mu_1 \nu_1 \dots \mu_{N} \nu_{N}} a^{\nu_1} \dots a^{\nu_N}=\psi^{\dagger}\Gamma^{(1)}_{\mu_1 \nu_1} a^{\nu_1} \dots \Gamma^{(N)}_{\mu_{N} \nu_{N}}a^{\nu_{N}}\psi 
\label{28}
\end{equation}
for any constant future-causal vector $a^{\mu}$, representing the four-velocity of an observer. The tensor $j^{\mu_1 \dots \mu_N}$ then obeys the conservation equation
\begin{equation}
\partial_0 j^{0_1 \dots 0_N} + \sum^N_{\alpha = 1}  \partial^{(\alpha)}_{i} j^{0_1 \dots i_{\alpha} \dots 0_N} = 0.
\label{29}
\end{equation}
Before giving a Bohmian model of the many-particle Kemmer theory, we show that $j^{0_1 \dots \mu_{\alpha} \dots 0_N}$ is future-causal for every $\alpha = 1, \dots ,N$. In section 2 it was shown that the operator $\Gamma^{(\alpha)}_{0 \mu} a^{\mu}$ is positive in spin space $\mathbb{C}^M$, where $M$ is the dimension of the representation of the $\beta$ martrices. As a result the operator
\begin{equation}
\Gamma= \Gamma^{(1)}_{0 \nu_1} a^{\nu_1} \dots \widehat{\Gamma^{(\alpha)}_{0 \nu_{\alpha}} a^{\nu_{\alpha}}} \dots \Gamma^{(N)}_{0 \nu_{N}}a^{\nu_{N}},
\label{30.00101}
\end{equation}
where the hat indicates that the term should be omitted from the product, is a positive operator in $N-1$ particle spin space $(\mathbb{C}^M )^{\otimes (N-1)}$. Consequently there exists an operator $\Omega$ in $N-1$ particle spin space such that $\Gamma = \Omega^{\dagger} \Omega$. As a result one can write
\begin{equation}
j_{0_1 \dots \mu_{\alpha} \dots 0_N} = (\Omega \psi)^{\dagger} \Gamma^{(\alpha)}_{\mu_{\alpha}\nu_{\alpha} } a^{\nu_{\alpha}}  (\Omega \psi).
\label{30.00102}
\end{equation}
This shows that $j_{0_1 \dots \mu_{\alpha} \dots 0_N}$ is the sum of $N-1$ (sum over all but the $\alpha^{th}$ spin index of $\Omega \psi$) vectors of the form $\Psi^{\dagger}\Gamma^{(\alpha)}_{\mu_{\alpha}\nu_{\alpha} } a^{\nu_{\alpha}} \Psi$. Because each such term is future-causal the sum $j_{0_1 \dots \mu_{\alpha}\dots 0_N}$ is also future-causal. In particular this implies that the energy density $j^{0_1 \dots 0_N}$ is positive and thus can be interpreted as a probability density.
  
It is now straightforward to construct a Bohmian model of the many-particle Kemmer theory. In this model, the description in terms of the Kemmer wavefunction is enlarged with $N$ particles which are guided by this wavefunction. The coordinates of the particles $x^{\mu}_n = (t_n,{\bf x}_n)$ (for $n=1 ,\dots, N$) are considered at equal times, i.e.\ $t_1= \dots = t_N = t$. The positive quantity $j^{0_1 \dots 0_N}$ is interpreted as the probability density of the Bohmian particles and the Bohmian guidance equation of the $\alpha^{th}$ particle can be defined as
\begin{equation}
\frac{\ud x^i_{\alpha}}{ \ud t} = \frac{j^{0_1 \dots i_{\alpha} \dots 0_N}}{j^{0_1 \dots 0_N}}.
\label{30}
\end{equation}
Because 
\begin{equation}
j^{0_1 \dots \mu_{\alpha} \dots 0_N} j_{0_1 \dots \mu_{\alpha} \dots 0_N} \geqslant 0
\label{30.001}
\end{equation}
for every $\alpha = 1, \dots ,N$, the motions of the Bohmian particles are again time--like or null. 

Note that equation ({\ref{30}}) is not covariant. However, at the expense of introducing a certain space-time foliation, it can be written in a manifestly covariant way \cite{durr99}. Future-causal four-vectors can then be constructed by contracting the current tensor $j^{\mu_1 \dots \mu_N}$ with additional future-causal vectors which arise from this space-time foliation. Nevertheless, we will not pursue this covariance in order not to distract the attention from the main objectives of this Letter. 

In the special case that $a^\mu = \delta^\mu_0$ the density of the Bohmian ensemble becomes $j^{0_1 \dots 0_N} = \psi^{\dagger} \psi$ and the guidance equations become
\begin{equation}
\frac{\ud x^i_{\alpha}}{ \ud t} = \frac{\psi^{\dagger} {\tilde \beta}^{(\alpha)}_i \psi}{\psi^{\dagger} \psi}.
\label{30.002}
\end{equation}
The resulting conservation equation ({\ref{29}}), which can also be directly derived from ({\ref{26.1}}), turns into 
\begin{equation}
\partial_0(\psi^{\dagger} \psi)  + \sum^N_{\alpha = 1}  \partial^{(\alpha)}_{i} (\psi^{\dagger} {\tilde \beta}^{(\alpha)}_i \psi)  = 0.
\label{30.1}
\end{equation}
In this case the Bohmian particle equations display a formal resemblance to the Bohmian equations for the multi-particle Dirac equation \cite{holland99,bohm5}.

As in section 2, one can now consider the possibility of adding a divergenceless tensor $A^{\mu_1 \dots \mu_{2N}}$ to the energy--momentum tensor $\Theta^{\mu_1 \dots \mu_{2N}}$ in equation ({\ref{27}}). This could possibly lead to different trajectory laws for the Bohmian particles. This time the condition that the energy density should be given by $\Theta^{\mu_1 \dots  \mu_{2N}} a_{\mu_1 } \dots a_{\mu_{2N} }$ for any observer with constant four-velocity $a^\mu$ implies that 
\begin{equation}
A^{\mu_1 \dots  \mu_{2N}} a_{\mu_1 } \dots a_{\mu_{2N} }= 0
\label{32}
\end{equation}
for every constant future-causal vector $a^{\mu}$. This in turn implies that the tensor $A^{\mu_1 \dots \mu_{2N}}$ should be antisymmetric in at least two indices. So, also in the many-particle case the energy--momentum tensor is not uniquely fixed by imposing a fixed energy density for every observer. Similarly as in the one-particle case we can further restrict the additional tensor $A^{\mu_1 \dots \mu_{2N}}$ by demanding that it should depend only on the multi-particle Kemmer wavefunction $\psi$ and its complex conjugate. This results in the conditions 
\begin{equation}
A^{0_1 \dots \mu_{\alpha} \dots 0_{2N}} = 0
\label{32.1}
\end{equation}
for every $\alpha$ which is odd. Contrary to the one-particle case these conditions do not in general imply that the full tensor $A^{\mu_1 \dots \mu_{2N}}$ should be zero. Hence there remains an indeterminacy on the multi-particle Kemmer energy--momentum tensor. This results in an indeterminacy in the vectors
\begin{equation}
{\bar j}^{0_1 \dots \mu_\alpha \dots 0_N} = j^{0_1 \dots \mu_\alpha \dots 0_N} + A^{0_1 \nu_1 \dots  \mu_\alpha \nu_\alpha   \dots 0_N \nu_{N}}  a_{\nu_1 } \dots a_{\nu_{N} }
\label{32.11}
\end{equation}
and hence in an indeterminacy in the defining equations for the Bohmian interpretation. Only in the Lorentz frame where the observer is at rest, i.e.\ $a^\mu = \delta^\mu_0$, the additional term containing the tensor $A^{\mu_1 \dots \mu_{2N}}$ vanishes in ${\bar j}^{0_1 \dots \mu_\alpha \dots 0_N}$ due to the conditions ({\ref{32.1}}). So in the frame where the observer is at rest, the Bohmian equations are unaffected by the remaining indeterminacy on the energy--momentum tensor. In a frame moving relative to the observer the additional tensor $A^{\mu_1 \dots \mu_{2N}}$ will in general affect the Bohmian equations. However the Bohmian equations defined in such a frame are physically less interesting because the Bohmian density $j^{0_1 \dots 0_N}$ is then not an observable quantity, only in a frame at rest relative to the observer, the Bohmian density $j^{0_1 \dots 0_N}$ is the energy density and is hence observable.

In summary we have established uniqueness of the Bohmian trajectories constructed from the energy--momentum tensor in the multi-particle case, under the assumptions that the energy--momentum tensor is conserved, that it gives the correct energy--momentum density for every observer with constant four-velocity and that it is only dependent on the Kemmer wavefunction and its complex conjugate. 

In the nonrelativistic limit, the Kemmer equations will result in the conventional many-particle Schr\"odinger equation in the spin-0 case. The corresponding Bohmian trajectory laws will result in those defined by Bohm for the many-particle Schr\"odinger equation:
\begin{equation}
\frac{\ud {\bf x}_{\alpha}}{ \ud t} = \frac{\textrm{Im} \big( \psi'^* {\boldsymbol \nabla}_{\alpha} \psi' \big)}{m |\psi'|^2}, 
\label{35}
\end{equation}
with $\psi'( {\bf x}_1 ,\dots , {\bf x}_N)$ the nonrelativistic $N$-particle wavefunction. In the nonrelativistic spin-1 case the Bohmian trajectory laws will contain an additional spin-dependent term similar as in equation ({\ref{25.05}}):   
\begin{equation}
\frac{\ud {\bf x}_{\alpha}}{ \ud t} =  \frac{\textrm{Im} ( \Phi^{\dagger}  {\boldsymbol \nabla}_{\alpha}  \Phi)}{m \Phi^{\dagger} \Phi} + \frac{{\boldsymbol \nabla} \times (\Phi^{\dagger} {\bf  {\hat  S}}_{\alpha} \Phi )}{2m\Phi^{\dagger} \Phi},
\label{36}
\end{equation}
with $ {\bf  {\hat  S}}_{\alpha}$ operating only on the $\alpha^{th}$ spin index of the nonrelativistic spin-1 multi-particle wavefunction $\Phi^{\alpha_1 ,\dots ,\alpha_N} ( {\bf x}_1 ,\dots , {\bf x}_N)$, where each index $\alpha$ runs from 1 to 3.

The multi-particle Kemmer charge tensor current is defined as 
\begin{equation}
s_{\mu_1 \dots \mu_N} = \psi^{\dagger}\eta^{(1)}_0 \beta^{(1)}_{\mu_1} \dots \eta^{(N)}_0 \beta^{(N)}_{\mu_N}  \psi
\label{37}
\end{equation}
and is conserved
\begin{equation}
\partial_0 s^{0_1 \dots 0_N} + \sum^N_{\alpha = 1}  \partial^{(\alpha)}_{i} s^{0_1 \dots i_{\alpha} \dots 0_N} = 0.
\label{38}
\end{equation}
Under similar conditions as for the Kemmer energy--momentum tensor one can prove that the vectors $s^{0_1 \dots \mu_\alpha \dots 0_N}$ are uniquely defined. Just as in the one-particle case the vectors $s^{0_1 \dots \mu_\alpha \dots 0_N}$ have the same nonrelativistic limit as the vectors $j^{0_1 \dots \mu_\alpha \dots 0_N}$.

\section{The minimally coupled Kemmer theory}
\subsection{Note on the minimally coupled Kemmer theory}
It is argued by Ghose {\em et al.} \cite{ghose93,ghose94,ghose96} that the program of constructing a conserved current from the energy--momentum tensor can be maintained when an interaction with an electromagnetic field $V^{\mu}$ is introduced via minimal coupling. However, in this section we will show that, when the minimal coupling is correctly introduced, the energy--momentum tensor $\Theta^{\mu \nu}_K$ is not generally conserved. This implies that the energy--momentum vector for an arbitrary observer $\Theta^{\mu \nu}_K a_\nu$ is not generally conserved. Remark that it is natural that the energy--momentum tensor (even for fermions) is not conserved when minimal coupling is introduced, because a charged particle can always exchange energy and momentum with the electromagnetic field. The charge current however, is always conserved for bosons and fermions because the electromagnetic field does not carry charge. Therefore the particle current in the Dirac case, which is the charge current, is conserved even in presence of an external electromagnetic field.

The discrepancy in the minimal coupled Kemmer theory, arises because there are two inequivalent ways of introducing it. One can introduce it at the level of the covariant form of the Kemmer theory ({\ref{1.1}}): \begin{equation}
(i\beta^{\mu} D_{\mu} - m )\psi = 0, 
\label{25.071}
\end{equation}
with $D_{\mu} = \partial_{\mu} + ie V_{\mu}$, or one can introduce it at the level of the Schr\"odinger form of the Kemmer theory ({\ref{13.1}}) and ({\ref{13.2}}), as proposed by Ghose {\em et al.}:
\begin{eqnarray}
i D_{0} \psi &=& (-i{\tilde \beta}_{i} D_{i} + m\beta_{0}   ) \psi \label{25.0711}\\
i \beta_{i} \beta^2_{0} D_{i}\psi &=& m (1 -  \beta^2_{0}) \psi.
\label{25.072} 
\end{eqnarray}
That these ways are not equivalent was already noted by Kemmer \cite{kemmer39}. The minimally coupled Schr\"odinger form implies the ``covariant form'', but not vice versa \cite{ghose96,kemmer39}. When the minimally coupled covariant form of the Kemmer equations is written in Schr\"odinger form, there appears an additional term in the Hamiltonian, i.e.\ equation ({\ref{25.0711}}) must be replaced by
\begin{equation}
i D_{0} \psi = (-i{\tilde \beta}_{i} D_{i} + m\beta_{0} ) \psi  - \frac{ie}{2m} F^{\mu \nu}(\beta_{\nu}\beta_{0}\beta_{\mu} + \beta_{\nu} g_{\mu 0})\psi,
\label{25.073}
\end{equation}
with $[D^{\mu},D^{\nu}] =ie F^{ \mu \nu }$. The additional term in the Hamiltonian, which is hard to interpret, has no equivalent in the spin-$\frac{1}{2}$ Dirac theory. It has recently been argued that this additional term is irrelevant \cite{nowakowski98,lunardi00}. The argument is that when the Kemmer theory is reduced to its physical components, then the Kemmer theory reduces to the minimally coupled Klein--Gordon theory in the spin-0 case and to the minimally coupled Proca theory in the spin-1 case where the anomalous term containing $F^{\mu \nu}$ disappears. However this doesn't settle the question whether or not we can safely introduce minimal coupling at the level of the Schr\"odinger form in the Kemmer theory. Because the minimally coupled Schr\"odinger form implies the minimally coupled covariant form, both the equations ({\ref{25.0711}}) and ({\ref{25.073}}) should be valid when the minimally coupled Schr\"odinger form is regarded as fundamental. Hence the additional term containing the tensor $F^{\mu \nu}$ should be zero. Without considering an explicit representation we can already show that this implies that the introduction of minimal coupling at the level of the Schr\"odinger form is in general untenable.

If one introduces minimal coupling at the level of the Schr\"odinger form of the Kemmer equation, one may derive from ({\ref{25.0711}}) (by multiplying ({\ref{25.0711}}) by $\psi^{\dagger}$ from the right, multiplying the conjugate of ({\ref{25.0711}}) by $\psi$ from the left, and substracting the two from each other \cite{ghose96}) that
\begin{equation}
\partial_0 (\psi^{\dagger}\psi ) +\partial_i (\psi^{\dagger}  {\tilde \beta}_i  \psi ) = 0
\label{25.9}
\end{equation}
or differently written $\partial_\mu \Theta^{\mu 0 }_K = 0$. If we assume that the the Schr\"odinger form of the Kemmer equation is Lorentz covariant, i.e.\ if we assume that the wave equation has the form ({\ref{25.0711}}) in every Lorentz frame, then one has $\partial_\mu \Theta^{\mu 0 }_K = 0$ in every Lorentz frame. Because $\partial_\mu \Theta^{\mu \nu }_K$ transforms as a four-vector under Lorentz transformations, this implies that the time component of $\partial_\mu \Theta^{\mu \nu }_K$ is zero in every Lorentz frame. Because the only four-vector which satifies this property is the zero vector, one has
\begin{equation}
\partial_\mu \Theta^{\mu \nu }_K = 0.
\label{34.021}
\end{equation}
Because the tensor $\Theta^{\mu \nu }_K$ is conserved and future-causal (the Kemmer energy--momentum tensor is unaltered by the introduction of an external field) one could continue the conventional program of constructing a conserved current which could be interpreted as a particle current \cite{ghose93,ghose94,ghose96}. However, if one departs from the minimally coupled Kemmer equation in covariant notation one can analogously derive from equation ({\ref{25.073}})
\begin{equation}
\partial_\mu \Theta^{\mu \nu }_K = e F^{ \nu \mu }s_{K\mu}.
\label{34.022}
\end{equation}
Because $s^{\mu}_K$ is the charge current, the right hand side of equation ({\ref{34.022}}) is the Lorentz force. Because the minimally coupled Schr\"odinger form implies the minimally coupled covariant form, both ({\ref{34.021}}) and ({\ref{34.022}}) should be valid when the minimaly coupled Schr\"odinger form is regarded as fundamental. This however implies that the Lorentz force is zero, i.e.\
\begin{equation}
F^{ \nu\mu }s_{K\mu} = 0.
\label{34.023}
\end{equation}
This is clearly a too severe constraint on the interaction field $V^{\mu}$ or on the wavefunction $\psi$. This suggests that we should introduce minimal coupling via the covariant form of the Kemmer equations. 

Consider now the specific case of spin-0. In the same way as was done in section 3, using the explicit representation, one can partially solve the Schr\"odinger form of the Kemmer equations. Both ways of introducing minimal coupling then lead to the minimally coupled Klein--Gordon theory. By using the explicit representation the Kemmer wavefunction can be written as  $\psi=( D_\mu \phi,m \phi)^T$ where $\phi$ satisfies the minimally coupled Klein-Gordon equation
\begin{equation}
(D_{\mu}D^{\mu} + m^2)\phi = 0,
\label{34.031}
\end{equation}
which may be derived from the Lagrangian
\begin{equation}
\mathcal{L} = (D_{\mu}\phi)^* (D^{\mu}\phi) - m^2 \phi^* \phi.
\label{34.032}
\end{equation}
The Kemmer energy--momentum tensor then reduces to
\begin{equation}
\Theta^{\mu \nu}_K =D^{\mu} \phi (D^{\nu} \phi)^* + (D^{\mu} \phi)^* D^{\nu} \phi - g^{\mu \nu} \big(D_{\alpha} \phi (D^{\alpha} \phi)^* - m^2 \phi^* \phi \big).
\label{34.04}
\end{equation}
This is in fact not the energy--momentum tensor arising from the Lagrangian ({\ref{34.032}}), see below. Equation ({\ref{34.023}}) reduces to $F^{ \nu\mu }s_{\! K\! G\mu} = 0 $ with $s^{\mu}_{KG}$ the charge current from the minimally coupled Klein--Gordon theory
\begin{equation}
s^{\mu}_{KG} =  i\big(\phi^* D^{\mu}\phi -\phi D^{\mu}\phi^*  \big) .
\label{34.041}
\end{equation}
The problem of vanishing Lorentz force thus doesn't disappear when the theory is reduced to its physical components. Because in the Klein--Gordon case only equation ({\ref{34.022}}) can be derived from the minimally coupled Klein--Gordon equation and not ({\ref{34.021}}), one should introduce minimal coupling at the level of the covariant form of the Kemmer equations, and not at the level of the Schr\"odinger form, in order to obtain the minimally coupled Klein--Gordon theory. However, when we introduce minimal coupling at the level of the covariant form of the Kemmer equations we in general loose the desired property that $\Theta^{\mu \nu}$ is conserved. 

The Klein--Gordon energy--momentum tensor which arises from the Lagrangian ({\ref{34.032}}) is neither conserved. This energy--momentum tensor, which reads
\begin{equation}
\Theta^{\mu \nu}_{KG} =D^{\mu} \phi \partial^{\nu} \phi^* + (D^{\mu} \phi)^* \partial^{\nu} \phi - g^{\mu \nu} \big(D_{\alpha} \phi (D^{\alpha} \phi)^* - m^2 \phi^* \phi \big),
\label{34.05}
\end{equation}
satisfies $\partial_{\mu}  \Theta^{\mu \nu}_{KG} = e (\partial^{\nu}V^{\mu}) s_{\! K\! G \mu}$. In addition the component $\Theta^{00}_{KG}$ of the energy--momentum tensor is only positive for a well defined class of potentials. 

The same remarks apply when we consider the spin-1 representation. So only in particular cases where the Lorentz force vanishes, for example in the case of an uncharged particle, the energy--momentum tensor is conserved and leads to a conserved energy--momentum vector which can be interpreted in the conventional way as a particle probability current. In other cases the energy--momentum tensor is not conserved. Although one can still maintain the notions of probability and flowlines in such a case, the non-conservedness implies that probability can be created and annihilated along the flowlines. A particle interpretation is hence untenable (simply because this would contravene with the current experimental knowledge), but the notion of energy flow, as introduced by Holland, could be maintained.

\subsection{Nonrelativistic limit of the minimally coupled Kemmer theory}
For the sake of completeness, we derive the nonrelativistic limits of the energy--momentum current and the charge current in the minimally coupled Kemmer theory. Remark that the energy--momentum tensor and charge current are unaltered by the introduction of an external field. Similar as in section 2, the uniqueness of the charge current is established by demanding that it reproduces the correct charge density in every frame. The uniqueness of the energy--momentum tensor can be established by demanding that it reproduces the correct energy--density in every frame, that it only depends on the wavefunction $\psi$, its complex conjugate $\psi^*$ and the interaction field $V^\mu$ and that it satisfies the conservation equation ({\ref{34.022}}). In the nonrelativistic limit the energy--momentum current and the charge current result in the same nonrelativistic current, in both the spin-0 and spin-1 representation. The nonrelativistic currents are conserved and have a positive time component and can thus be interpreted as particle probability currents. Hence in the nonrelativistic limit a Bohmian particle interpretation of these currents is possible. If we demand that the nonrelativistic particle currents should be derivable from the charge current or energy--momentum current, then we obtain a unique (under the specified conditions) nonrelativistic Bohmian particle interpretation for spin-0 and spin-1 in the coupled case.  

In the nonrelativistic limit the minimally coupled Klein--Gordon equation ({\ref{34.031}}) becomes the minimally coupled Schr\"odinger equation
\begin{equation}
i\frac{\partial \psi'}{\partial t} = -\frac{({\boldsymbol \nabla} - ie {\bf V})^2 \psi'}{2m} + e V_0 \psi'.
\label{40}
\end{equation}
The nonrelativisic limit of both the Kemmer energy--momentum vector $\Theta^{\mu \nu}_K a_\nu$ and the Kemmer charge current $s^\mu$ become the conserved particle probability current associated with the minimally coupled Schr\"odinger equation
\begin{equation}
j^0  \! = |\psi'|^2 , \quad {\bf j} = \frac{1}{m} \textrm{Im} ( \psi'^* ({\boldsymbol \nabla} - ie {\bf V})  \psi' ).
\label{41}
\end{equation}
Hence the corresponding Bohmian particle interpretation is the conventional one \cite{holland99,bohm5}. 

When the spin-1 representation of the minimally coupled Kemmer theory is reduced to its physical components, it results in the minimally coupled Proca theory which can be obtained from the free Proca theory by replacing the derivative $\partial^\mu$ by the covariant derivative $D^\mu = \partial^\mu + ie V^\mu$. In the nonrelativistic limit the minimally coupled Proca equation becomes the nonrelativistic spin-1 equation for the three-component wavefunction $\Phi$
\begin{equation}
i\frac{\partial \Phi}{\partial t} = -\frac{ ({\boldsymbol \nabla} - ie {\bf V})^2 \Phi}{2m}  + e V_0 \Phi + \frac{e {\bf B} \cdot {\bf {\hat S}} \Phi }{2m}.\label{42}
\end{equation}  
where ${\bf B}$ is the magnetic field. The nonrelativistic limit of the currents $\Theta^{\mu \nu}_K a_\nu$ and $s^\mu$ is
\begin{equation}
j^0 = \Phi^{\dagger}  \Phi, \quad {\bf j} = \frac{1}{m} \textrm{Im} ( \Phi^{\dagger}  ({\boldsymbol \nabla} - ie {\bf V} )  \Phi) + \frac{1}{2m} {\boldsymbol \nabla} \times (\Phi^{\dagger} {\bf  {\hat  S}} \Phi ).
\label{43}
\end{equation}
This current differs from the conventional one by the spin-dependent term. Hence if the nonrelativistic current should be derivable from the relativistic charge current or energy--momentum current, then the nonrelativistic Bohmian guidance equation should contain an additional spin term in the spin-1 case.

\section{Conclusion}
We established the uniqueness of the Bohmian particle interpretation constructed from the conserved energy--momentum tensor in the case of spin-0 and spin-1 particles, under well determined conditions. Similar techniques could be used to show that the massless spin-0 and spin-1 theory, i.e.\ the Maxwell theory, only admits one possible construction of causal tracks from the conserved energy--momentum tensor. To give a Bohmian inspired trajectory interpretation similar to that of the Dirac or Kemmer case, one has to start from the Harish-Chandra formalism \cite{ghose93,ghose96,ghose01,harish46}, which introduces minor modifications to the Kemmer formalism to describe massless spin-0 or spin-1 particles. The uniqueness of the causal trajectories then follows from the uniqueness of the energy--momentum tensor of the electromagnetic field, which was establised by Fock \cite{fock64}.

\vspace*{6mm}
\noindent
Acknowledgements:

\vspace*{6mm}
\noindent
W.S.\ acknowledges financial support from the F.W.O.\ Belgium. J.D.N.\ acknowledges financial support from IWT (``Flemish Insitute for the scientific-technological research in industry''). W.S.\ is also grateful to Peter Holland for helpful comments and to Partha Ghose for useful discussions.

\end{document}